\documentclass{aa}
\usepackage{graphicx}
\usepackage{txfonts} 
\usepackage[utf8]{inputenc} 
\usepackage[dvipsnames]{xcolor}
\begin{document} 

\authorrunning{Locci et al.}
\titlerunning{The evolution of close-in gas giants}
\title{Photo-evaporation of close-in gas giants \\ orbiting around G and M stars}
\author{Daniele Locci, Cesare Cecchi-Pestellini and Giuseppina Micela}
\institute{INAF ~\textendash~ Osservatorio Astronomico di Palermo, Piazza del Parlamento 1, I-90134 Palermo, Italy \\ \email{dlocci@astropa.unipa.it}}
\date{Received; accepted}

\abstract
{X-rays and extreme ultraviolet radiation impacting on a gas produce a variety of effects that, depending on the electron content, may provide a significant heating of the illuminated region. In a planetary atmosphere of solar composition, stellar high energy radiation may heat the gas to very high temperatures, that may have consequences on the stability of planetary atmospheres, in particular for close-in planets.}
{We investigate the variations with stellar age in the occurring frequency of gas giant planets orbiting G and M stars, taking into account that the high energy luminosity of a low mass star evolves in time, both in intensity and hardness.}
{Using the energy-limited escape approach we investigate the effects induced by the atmospheric mass loss on giant exoplanet distribution that is initially flat, at several distances from the parent star. We follow the dynamical evolution of the planet atmosphere, tracking the departures from the initial profile due to the atmospheric escape, until it reaches the final mass-radius configuration.} 
{We find that a significant fraction of low mass Jupiter-like planets orbiting with periods lower than $\sim 3.5$ days either vaporize during the first billion years, or lose a relevant part of their atmospheres. The  planetary  initial  mass  profile  is  significantly  distorted; in particular,  the frequency of occurrence of gas giants, less massive than $2~M_J$, around young star can be considerably greater than the one around older stellar counterparts.}
{}
\keywords{Planets and satellites: gaseous planets -- Planets and satellites: atmospheres -- Planets and satellites: dynamical evolution and stability -- Planet-star interactions}
\maketitle

\section{Introduction}
X-rays and extreme ultraviolet radiation impacting on a gas produce a variety of effects, that depending on the electron content, may provide a significant heating of the illuminated region. In a planetary atmosphere of solar composition X-rays penetrate much deeper than ultraviolet radiation (e.g. \citealt{Cp09}), where due to the large fractional ionization  may heat the gas to very high temperatures, having consequences on the stability of planetary atmospheres, in particular for close-in planets. 

After the dispersal of the protoplanetary disk, extrasolar planets generally cool and contract to their currently observed sizes. In planets close enough to the host star, the atmosphere may instead go through a long phase of efficient hydrodynamic escape, or even blow-off, with the atmosphere final fate depending mainly on the atmospheric mass, and the stellar irradiation (e.g. \citealt{L03}). The transiting exoplanet HD~209458b has been the first planet with an observed on going atmospheric hydrodynamic escape, estimated to be $\sim 10^{10}$~g~s$^{-1}$ \citep{VM03}. Few other planets with escaping atmospheres have been subsequently discovered, through (mainly) Ly$\alpha$ transit spectroscopy, as the case of the hot Neptune GJ 436b \citep{K14,Eh15,La17}, and the hot Jupiter HD 189733b \citep{Lec10,Bou13}.

At very early phases of planetary evolution, atmospheres are subjected to extremely intense high-energy stellar fluxes beyond the Lyman continuum edge (hereafter XUV radiation). This is the time when atmospheric escape is strongest and may shape the planetary envelope, before setting it onto its final evolutionary path. In fact, photo-evaporation of low-density atmospheres may be the dominant evolutionary mechanism in the planet size distribution \citep{FP18}. There is less of a consensus whether planetary evaporation is driven by extreme ultraviolet or X-ray heating. \citet{OJ12} pointed out that these two energy sources may give rise to separate effects. Evaporation driven by X-rays generally occurs at high X-ray luminosities, low planetary densities, high planetary masses, and small separations, while extreme ultraviolet seems dominate at low X-ray luminosities, high planetary densities, low planetary masses and large separations. These authors also found that the evaporative flow may undergo a transition from to be X-ray driven at early times to extreme ultraviolet driven at late times.

 To recover a planetary evolutionary path, we may start from the current mass of the planet, and retrace its evolution back in time (e.g., \citealt{Lo12} in the case of low mass planets), obtaining the mass of the planet when the star was much younger. On the other side taking the opposite view, we may consider a sample of giant planets with a uniform distribution in mass, assigning to each of them a radius from existing numerical calculations, at the initial time of the simulation, and follow the resulting evolution. 

The general characteristics of the mass loss process may be studied in a statistical sense, without modelling in detail individual planets using a method put forward by \citet{Lec07}. Through this approach consisting in the comparison of the stellar high energy received  by the an upper atmosphere to the planetary potential energy in an energy diagram, several studies (e.g., \citealt{Lec07,DeW09,EheD11}) discuss the effects of the mass loss on the evolutionary history of a population of known exoplanets.

Among the methods proposed to describe the hydrodynamical instability of a planetary atmosphere exposed to strong high-energy irradiation, the concept of energy-limited escape has been widely used in literature (e.g.  \citealt{E07,SF10,LoF13,Jin14}). In such a model put forward by \citet{W81}, the planetary thermosphere is considered as a closed system, and the mass-loss rate is derived by equating the radiative energy input to the energy gain of the evaporated atmospheric material. By using the energy-limited escape description, we investigate the effects in time of the atmospheric mass loss on a synthetic sample of giant exoplanets with  an initial uniform (flat) planetary mass distribution at several distances from a parent star. We extend the studies of \citet{PML08} and \citet{PM08}, who addressed such a problem exploiting synthetic planetary populations around a dG and dM stars, respectively. We  take into account the variations of the planetary size with time, in response to gravitational contraction and mass loss processes. Moreover, we assign to each planet a specified stellar luminosity evolving with time.

We assume an initial time of 10~Myr, when planets are supposed to have already reached their final orbit through migration (we  neglect type~III migration). Then, we follow the dynamical evolution of the planet atmosphere, tracking the departures from the initial profile due to the atmospheric escape, until it reaches the final mass-radius configuration. On this basis we may derive the fractional occurrence frequency of giants planets around young dwarf stars. Recent advances in detection techniques allow to reveal planets even around active young T-Tauri stars \citep{D17}, with the result that a larger incidence of hot Jupiter seems to occur at young stellar ages than in older phases. In our study we therefore investigate gas giant frequency variations at different orbital periods, in going from the T~Tauri phase to more evolved main sequence stages. Planetary periods are taken from the work of \citet{F13}, which used a catalog of more than 2300 candidate transiting planets, released in February 2013 by the mission Kepler \citep{Bat13}. These authors take into account five classes of exoplanets with orbital periods in the range $0.8-418$~days. Such distribution does not depend on the spectral type of parent stars, and it is valid from F to M spectral types. 

In the  following Section we present an outline of the method together with a description of computational approximations and shortcuts. Section \ref{res} contains a collection of results, while the conclusions we reach are presented in the last Section.

\section{The model}
\label{model}
We study a  population of synthetic giant planets orbiting around dG and dM stars. We initially assume a uniform mass distribution for a population of ${\cal N}_{\rm P}$ giant planets with masses, $M_{\rm P}$ in the range $0.2-16 \, M_J$, $M_J$ being the Jupiter mass. We subdivide the mass interval in twenty, evenly-spaced mass bins. We also consider the eleven orbital period bins selected by \citet{F13}. At the initial time, $t_{\rm i} = 10$~Myr, planet radii, $R_{\rm P}$ are taken from \citet{F07}, who derived the radius as a function of time given a planetary mass in the range $M_{\rm P} = 0.24-11.3~M_J$, and a planet mean orbital distance, $d_{\rm P}$ ranging from 0.02 to 9.5~AU. Planetary radii corresponding to masses laying beyond the limits have been extrapolated. For any mass bin, we choose 200 evenly spaced masses; then, for a given planetary mass we randomly assign one orbital period in each interval provided by \citet{F13}. The number of selected planets adds to a total ${\cal N}_{\rm P} = 44,000$. 

\begin{figure}
\centering
\includegraphics[width=.5\textwidth]{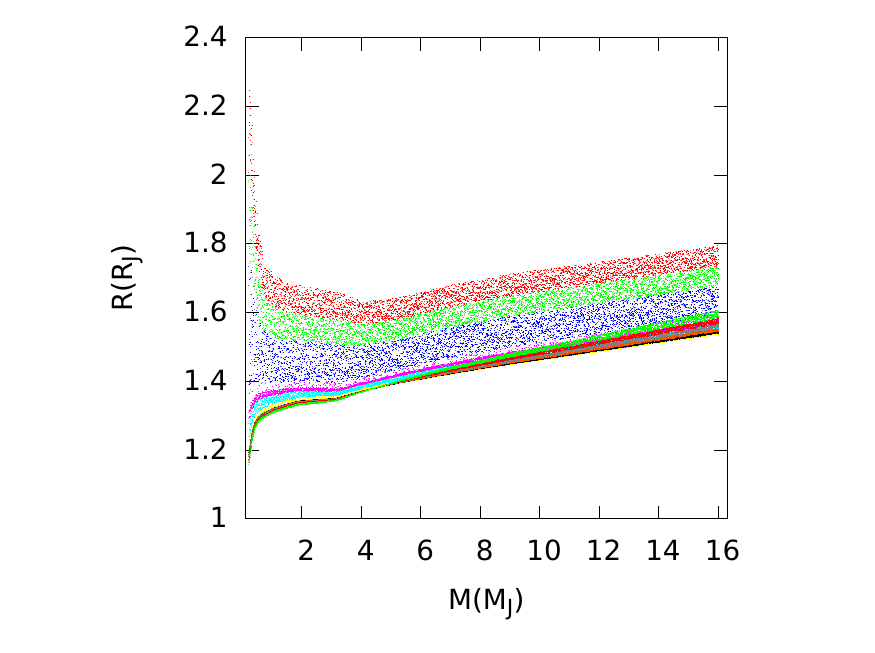}
\caption{Initial $M_{\rm P}-R_{\rm P}$ distribution at 10 Myr according to \citet{F07}, for a sample of ${\cal N}_{\rm P} = 44,000$ planets. Different orbital periods are shown in different colors. The periods bin are: $0.8-2$, $2-3.4$, $3.4-5.9$, $5.9-10$, $10-17$, $17-29$, $29-50$, $50-87$, $87-145$, $145-245$, $245-418$~days.  Periods grow in going from the upper to the lower parts of the diagram.}
\label{fig1}
\end{figure}

The model put forward by \citet{F07} have been developed for solar-like stars. Nevertheless, we extend it to cover stellar types up to M, scaling the stellar flux accordingly to the different orbital distances owned by planets with identical orbital periods revolving around different stars. In order to chose a suitable Fortney's track for the case of dM stars, we opportunely scale the distance from the star by a factor $\sqrt{L_{\odot}/L_{\rm M}}$, where $L_{\odot}$ and $L_{\rm M}$ are the luminosity of the dG and the dM stars respectively. Figure~\ref{fig1} shows the initial $M_{\rm P}-R_{\rm P}$ distribution of the ${\cal N}_{\rm P}$ model planets. Close-in planets are located in the upper part of the figure.

In the energy limited escape approach, the thermal planetary mass-loss rate results (e.g. \citealt{SF11})
\begin{equation}
\label{eq:maslos3}
\frac{{\rm d} M}{{\rm d} t} = \varepsilon \frac{\pi R^3_{\rm P} F_{\rm XUV}}{G M_{\rm P} K}
\end{equation}
where $F_{\rm XUV}$ is XUV radiation flux at the planet orbit, $G$ the gravitational constant, $K$ the potential energy reduction factor due to stellar tidal forces \citep{E07}, and $\varepsilon$ the escape efficiency. This latter quantity incorporates the details of the escape process, representing the efficiency of conversion of the energy of the incident XUV radiation into effective escape of gas. \citet{S07} consider such an efficiency as the product of the heating efficiency and the fraction of the deposited energy that is eventually lost through escaping gas. While this latter fraction has been estimated to be around 0.5, or lower (e.g. \citealt{P08}), the heating efficiency depends on the spectral shape of the incoming radiation \citep{Cp09}.

The high energy luminosity of solar type stars evolves in time, both in intensity and hardness (e.g. \citealt{M02}). We define an average heating efficiency weighted over the shape of a chosen stellar spectrum as follows
\begin{equation}
\varepsilon_{\rm h}(t) = \frac {\int_{\rm XUV} \eta(E) {\cal S} (E,t)~{\rm d} E} {\int_{\rm XUV} {\cal S} (E,t)~{\rm d} E}
\end{equation}
where $\eta$ is the heating efficiency for a hydrogen-ionizing photon of energy $E$ \citep{D99}, and ${\cal S}$ the spectral shape \citep{L18}. The variation in time of the spectral shape is estimated as 
\begin{equation}
{\cal S} (E,t) = c_S(t) {\cal S}_S (E) + c_H(t) {\cal S}_H (E)
\end{equation}
where $c_S$ and $c_H$, such that $c_S(t) + c_H(t) = 1$, are evolving coefficients taking into account the different evolution of the shapes of the emitted spectrum in the soft and hard XUV spectral bands (see \citealt{M02} for details). The adopted spectral shapes are models for the thermal emission of hot plasmas with energies 0.5 (soft band $S$) and 1~keV (hard band $H$), derived by \citet{RS77}. We obtain $\varepsilon (t) = 0.5 \times \varepsilon_{\rm h}(t) \la 0.4$.

Young solar type stars emit X-rays at a level $3-4$ orders of magnitude higher than the present-day Sun, during both the pre-main sequence phase when the emission is dominated by intense flares, and the first phases of the main sequence \citep{F03,Fa05}. In order to account for this effect in the X-ray band, we use the prescriptions given in \citet{PML08} and \citet{PM08}. We use as reference the stars in the Pleiades, a widely studied young  stellar cluster, with an estimate age of 80~Myr. The evolution of  the X-ray luminosity results
\begin{equation}
\label{lxtime}
\left(\frac{{L_{\rm X}}} {L^{(0)}_{\rm X}} \right) = a \times \left( \frac{\rm 1~Gyr}{\tau} \right)^b 
\end{equation}
where $L^{(0)}_{\rm X}$ is the initial luminosity of the star, and $\tau$ the star age in Gyr. The parameters $a$ and $b$ in equation~(\ref{lxtime}) are reported in Table~\ref{tone}.

\begin{table}
\centering
\caption{Parameters describing the evolution of X-ray luminosities of dG and dM stars, equation~(\ref{lxtime}).}
\scriptsize
\begin{tabular}{ccccccc}
\hline 
star & $L^{(0)}_{\rm X}$/erg~s$^{-1\dag}$ &  & \multicolumn{2}{c}{$a$} & \multicolumn{2}{c}{$b$} \\ 
\cline{4-7}
& & $(\tau/\rm 1~Gyr)$ & $\leq 0.6$ & $> 0.6$ & $\leq 0.6$ & $> 0.6$ \\ 
dG$^{(2)}$ & $10^{29.35}$ & & 0.379 & 0.19 & 0.425 & 1.69 \\
dM$^{(3)}$ & $10^{28.75}$ & & 0.17 & 0.13 & 0.77 & 1.34 \\
\hline 
\end{tabular}
\flushleft
$\dag$ \citet{PF05}; $(2)$ \citet{PML08}; $(3)$ \citet{PM08}.
\label{tone}
\end{table}

The X-ray luminosity of a star at specific stellar ages follows a distribution, whose  mean value is determined by relation~(\ref{lxtime}), with a spread of about one order of magnitude. This probability density function may be represented by a log-normal profile, whose cumulative distribution function reads as \citep{PML08}
\begin{equation}
\label{cdf}
{\cal P}[{\rm ln} (L_{\rm X}/{\rm erg~s^{-1}})] = \frac{1}{2} \left[ 1+ {\rm erf} \biggl( \frac{{\rm ln}(L_{\rm X}/{\rm erg~s^{-1}})-\mu}{\sqrt{2}\sigma} \biggr) \right]
\end{equation}
where $\mu = {\rm ln} (L^{(0)}_{\rm X}/\rm erg~s^{-1})$,  and $\sigma$ is the standard deviation of the logarithm variable \citep{PM08,PML08}. The adopted values for the initial X-ray luminosities are $L^{(0)}_X = 10^{29.35}$ and $10^{28.75}$~erg~s$^{-1}$ for dG and dM stars, respectively \citep{PF05}.

We consider ${\cal N}_{\rm X} = 5000$ X-ray luminosity bins, and we derive the fractional number of stars falling within each bin exploiting equation (\ref{cdf}). Then, for each of the 11 planet orbital bins, we randomly assign 4,000 X-ray luminosity (corresponding to the number of planets in each period bin), and follow their evolution in time through equation~(\ref{lxtime}). 

The extreme ultraviolet luminosity is given by the relation \citep{SF11} 
\begin{equation}
{\rm log} (L_{\rm EUV}/{\rm erg~s^{-1}})= 4.8 + 0.86 \times {\rm log} (L_{\rm X}/{\rm erg~s^{-1}})
\label{EUV}
\end{equation}
where it is assumed that the evolution of the extreme ultraviolet radiation follows that of the X-ray band. The total luminosity $L_{\rm XUV}$ is then the superposition of the luminosities in the  extreme ultraviolet and X-ray spectral bands.

The stellar irradiation opposes to the natural contraction of  the radius of a giant planet under his own weight. The rate at which the radius decreases is determined by the bond albedo, the presence of clouds and their optical depths, the gas-phase abundances and their opacities, and the intensity of the stellar radiation \citep{Bur00}. Isolated giant planets or brown dwarfs contract more rapidly than irradiated close-in planets (e.g. \citealt{Burr97}). For this reason the mass-radius depends on the orbital distance. In our simulation the radius is then allowed to change with time in response to the mass loss induced by the XUV radiation, together with the natural contraction of the planetary atmosphere due to the radiative losses. 

We let the planet to lose mass at time $t$ following the overflow from the Roche lobe induced by XUV radiation, and then we derive the radius (and therefore the density) corresponding to the new mass at the time $t+{\rm d}t$, scaling appropriately along the \citet{F07} tracks. We iterate until we reach the final time $t_{\rm f}=4.5$~Gyr.

\section{Results} \label{res}
We consider the cases of a dG star with mass $M_\star = 1~M_\odot$, and a dM star with $M_\star = 0.3~M_\odot$. Unless otherwise stated, each planet of mass $M_{\rm P}$ is characterized by two random parameters (simulated as explained above), namely the stellar X-ray luminosity $L_{\rm X}$, and the mean orbital period $T_{\rm P}$ (or the orbital distance $d_{\rm P}$).  

\subsection{Energy-diagrams}
Following \citet{Lec07}, we construct energy diagrams, in which the potential energy, $E_{\rm p}$ of a planet is plotted versus the power received by its upper atmosphere, ${\rm d}E_{\rm XUV}/{\rm d}t$. Such an approach allows to estimate the planetary lifetime and the atmospheric mass loss rate, $\dot{M}$ (g~s$^{-1}$).

\begin{figure*}
\centering
\includegraphics[width=0.75\textwidth]{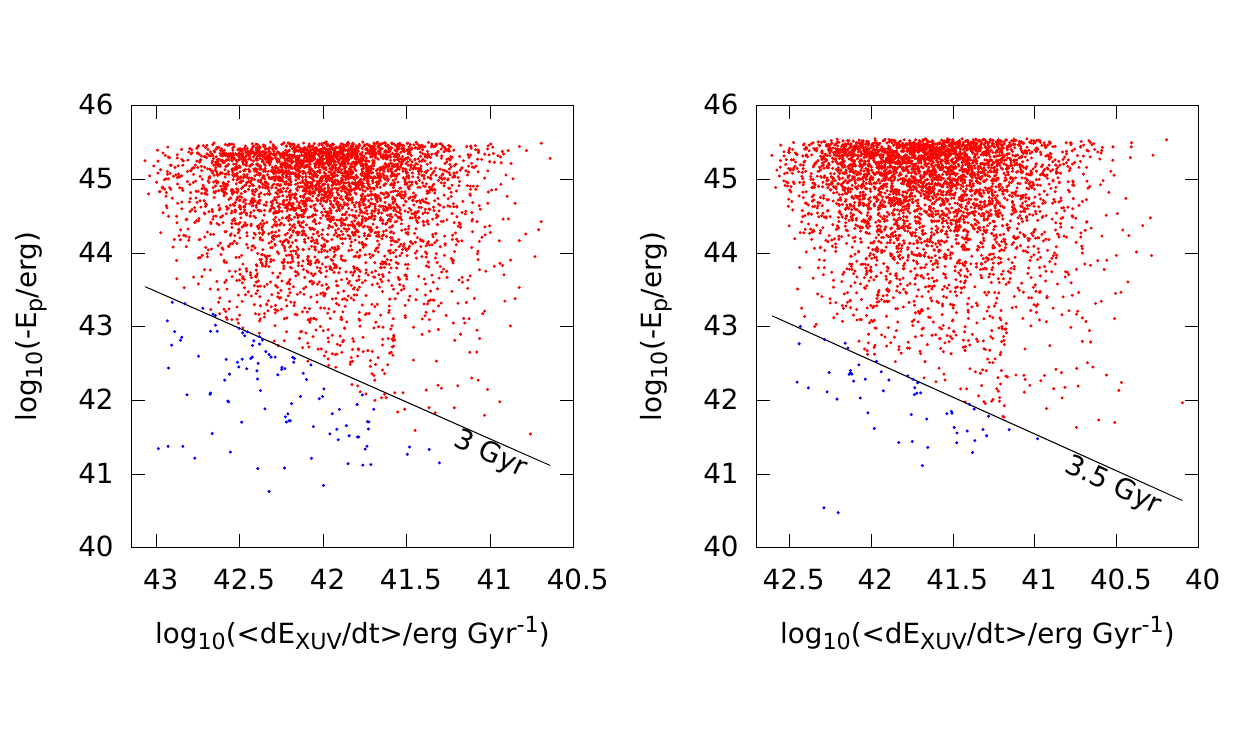}  
\caption{The energy diagram for planets orbiting dG (left panel) and dM (right panel) stars. Red dots represent planets surviving with some fraction of the initial envelope, while blue dots  planets that lost the entire envelope. The solid black line marks the lifetime line.}
\label{fig2}
\end{figure*}

In Figure \ref{fig2} we show the energy diagrams for our set of ${\cal N}_{\rm P} (= 44,000)$ models, orbiting around dG and dM stars. All the planets that have lost completely their envelope lie in the evaporation-forbidden region delimited by the lifetime line. 

\begin{figure*}
\centering
\begin{tabular}{c}
\scriptsize
\includegraphics[width=.95\textwidth]{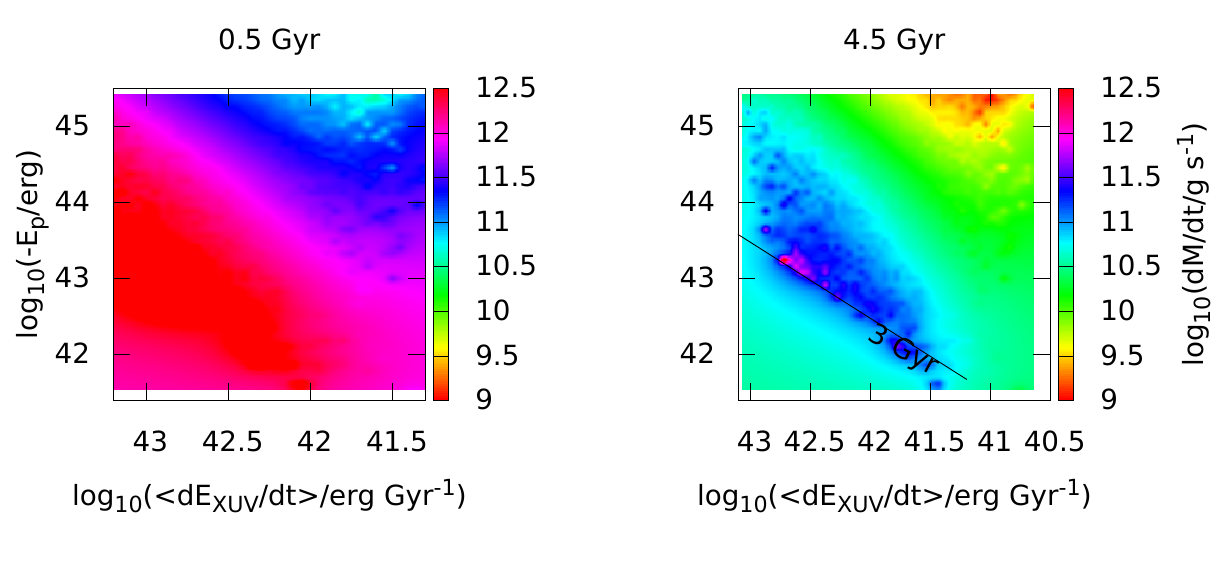} \\ 
\includegraphics[width=.95\textwidth]{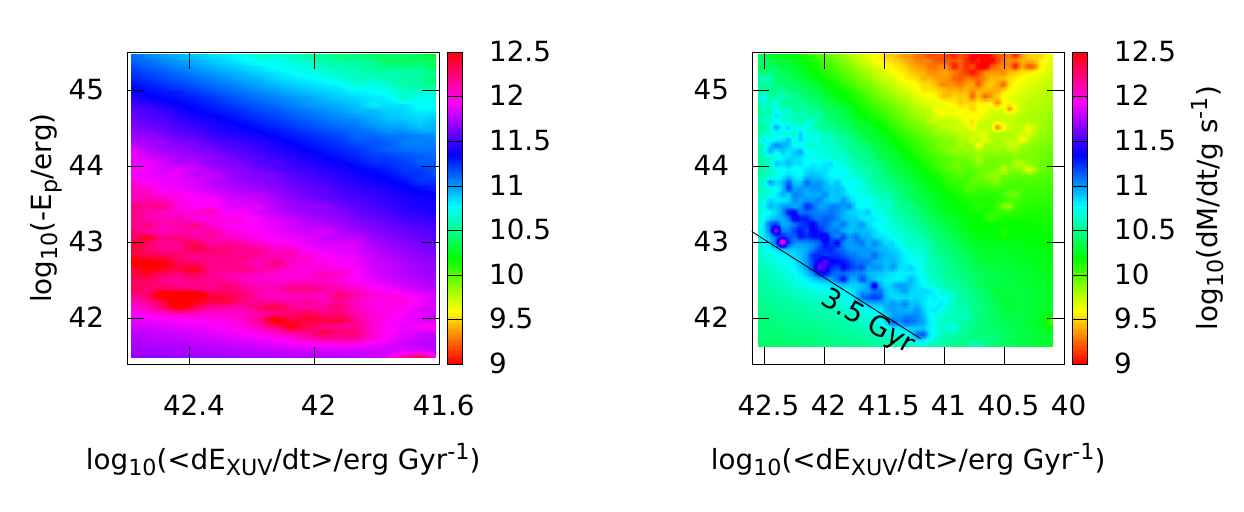} \\
\end{tabular}
\caption{Color map of the mass loss rate at 0.5 and 4.5 Gyrs in planets orbiting dG (top panels) and dM (bottom panel) stars.}
\label{fig3}
\end{figure*}

The lifetime in the case of dM stars is slightly greater than the case of dG stars, implying that planets survive longer around dM stars, as shown in Figure~\ref{fig3}. The mass loss rate has been calculated by means  equation (\ref{eq:maslos3}), through the model described in Section~\ref{model}. The ratio of the stellar fluxes received by planets orbiting around dG and dM stars at the same orbital distance is $F_\odot /F_{\rm M } = L_\odot M_{\rm M}/ L_{\rm M} M_\odot \sim 1.8$.

\subsection{Planetary mass distribution}
The main objective of this work is to study the effects of the high energy stellar radiation on planetary mass distribution.  Distant planets are poorly subjected to mass loss events, their radii being mainly constrained by natural contraction, while closer planets, and among them, those with lower densities may be significantly affected.

\begin{figure*}
\centering
\begin{tabular}{c}
\includegraphics[width=.8\textwidth]{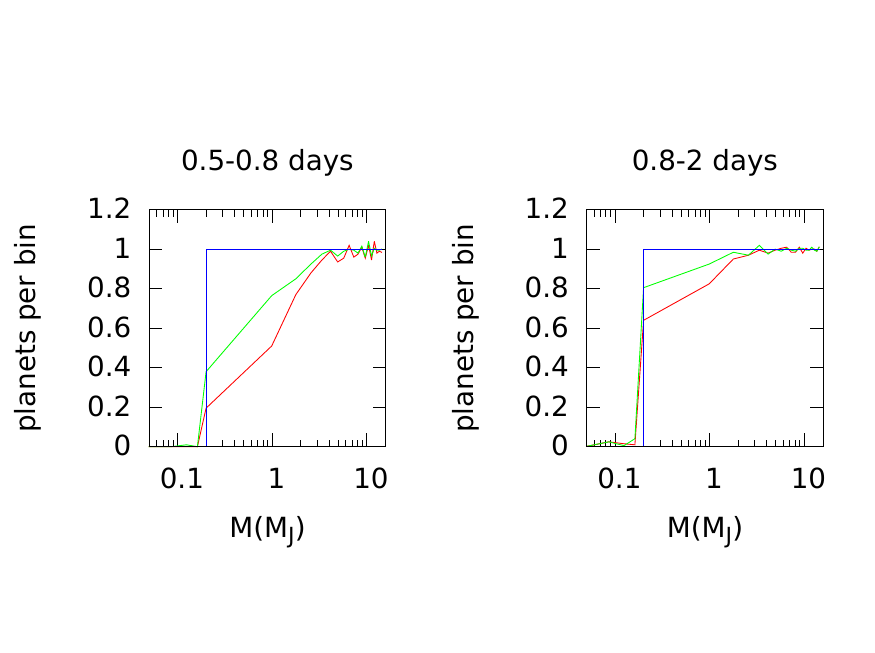} \\
\includegraphics[width=.8\textwidth]{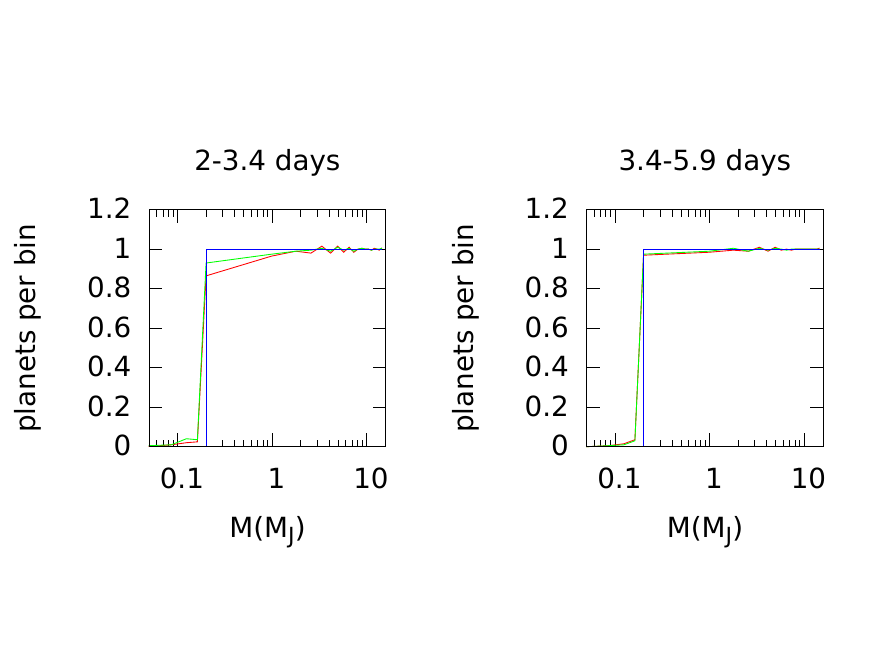}
\end{tabular}
\caption{Normalized number of planets per mass bin. Each panel corresponds to an orbital period interval. Initial (flat) mass distribution: blue line; final mass distribution for planets around dG stars: red line; final mass distribution for planet around dM stars: green line. Since radiative effects are negligible, orbital periods longer than $\sim 6$~days have not been shown.} 
\label{fig4}
\end{figure*}

In Figure~\ref{fig4} we show the effects of XUV irradiation over the initial, flat mass distribution of planets orbiting either dG and dM stars for several period bins. The number of planets of each mass bin is normalized at the number of planets for bin of the initial flat mass distribution. We find that for periods longer than 6 days, the effects of high energy radiation are  negligible. Low mass Jupiter-like planets close to the star are instead strongly perturbed. A large fraction of of planets with $M_{\rm P} \la 0.6~M_J$ located very close to their parent stars (first Fressin's period bin) are vaporized. The most affected planets are those with a large radius (low density), orbiting around X-ray bright stars. Over the entire planetary mass distribution the percentage of planets that are lost, i.e. with final masses outside the range $0.2 - 13~M_J$ never exceeds 4\% (2\%) for dG (dM) stars in the first period bin ($0.8-2$ days). However, the number of lost planets is not evenly distributed over the entire range of masses, but is concentrated at lower masses. At the shortest assumed orbital periods $(T_{\rm P} = 0.8-2$~days), 36\% (20\%) of planets in the $0.2-1~M_J$ mass bin, around dG (dM) stars is lost, and thus removed from the initial distribution; the same fate occurs to 18\% (8\%) of planets falling in the $1-2~M_J$ mass bin. The small bump located at low mass values in the mass profile shown in Figure~\ref{fig4} is produced by planets falling outside the lower boundary of the original mass distribution, who still maintain residual atmospheres. The cores of vaporized planets are not shown in the Figure~\ref{fig4}. 

In order to investigate the effects of extreme ionizing fluxes, we consider model planets orbiting with rather short periods, i.e. $T_{\rm P} = 0.5-0.8$~days. In the left-top panel of Figure~\ref{fig4} is reported the final mass distribution. In the case of dG (dM) stars, 11\%  (\%7) of planets are lost in the full mass range ($0.2-13 M_J$), and lose completely their atmosphere in $\sim 9$\% ($\sim 6$\%) of the total number of planets. We also find that 81\% (62\%) is removed from the mass range $0.2-1~M_J$, while 49\% (24\%) is removed from the $1-2~M_J$ mass bin.  Of the original 200 planets initially present in the first mass bin $0.2-1~M_J$, 49.5\% (8\%) of them are removed as they are inside the Roche limit,  49\% (78\%) are vaporized, while the remaining 0.5\% (14\%) survive. These planets are not of course the solely population of the  $0.2-1~M_J$ mass bin, as this small fraction of planets is increased by the overflow from higher mass bins.  

\begin{figure*}
\centering
\begin{tabular}{cc}
\scriptsize
\includegraphics[width=.5\textwidth]{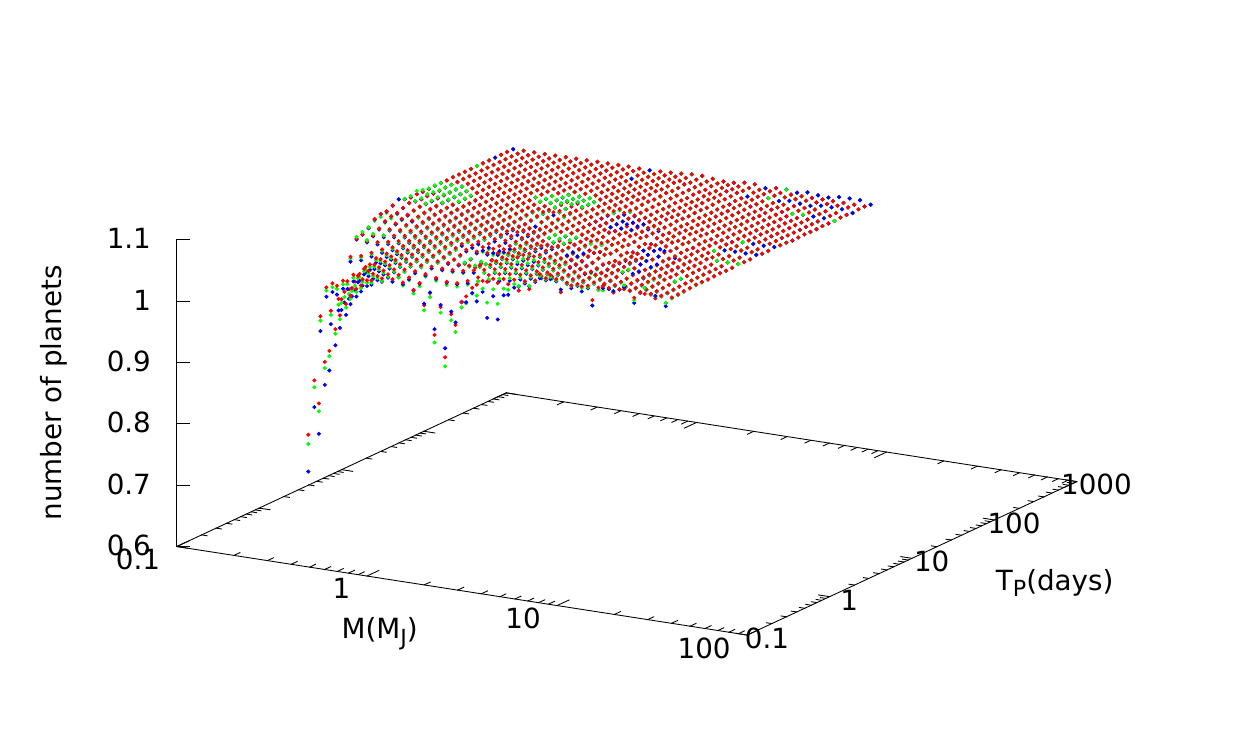}  & \includegraphics[width=.5\textwidth]{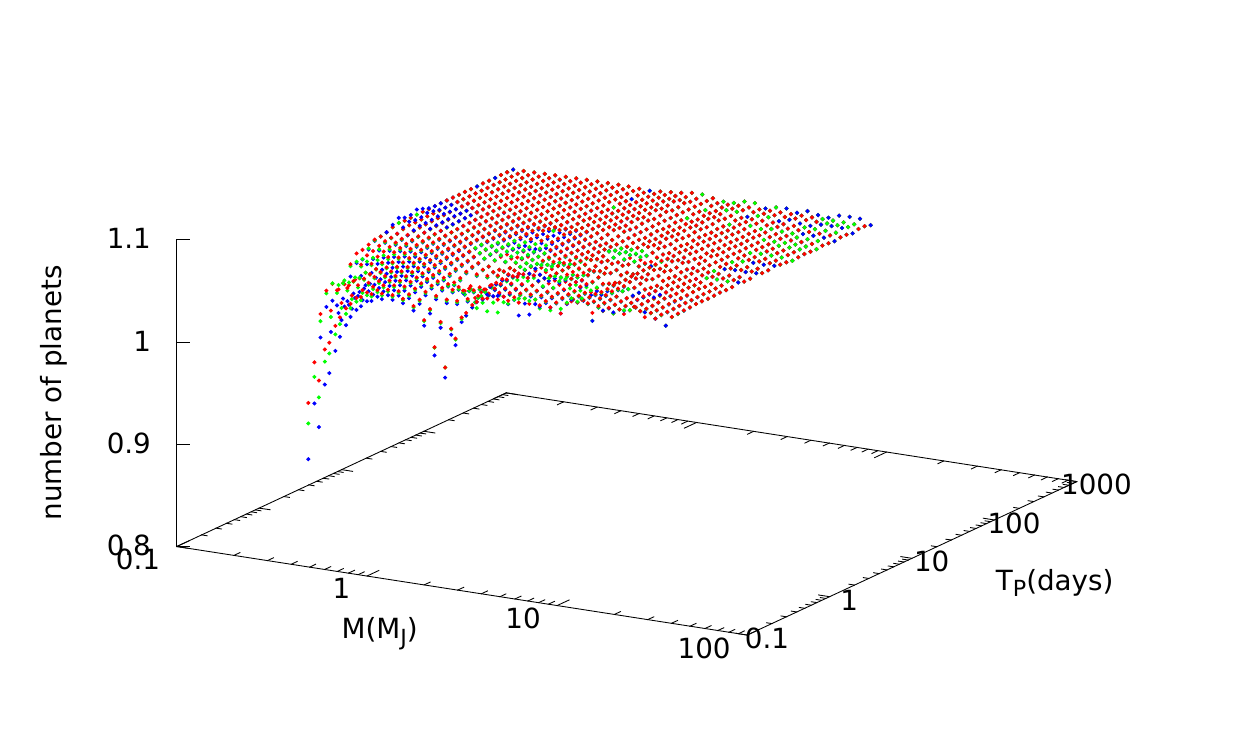}
\end{tabular}
\caption{Normalized number of planets as function of the mass and of the orbital period; the plot summarizes the data for three different ages, 0.5 (red dots), 1.5 (green dots) and 4.5 (blue dots) Gyr.}  
\label{fig5}
\end{figure*}

In Figure~\ref{fig5} we include all orbital periods, and we show the behavior of the mass distribution for both dG and dM stars. For the sake of simplicity we include three ages, 0.5, 1.5, and  4.5 Gyr, adding to a total of $44,000 \times 3$  model points. As expected the more affected planets are those with low mass and small orbital periods. This is consistent with e.g., \citet{SeK11} findings that described the so called  evaporation sub-Jupiter desert in the range mass of $0.02-0.8 M_J$. For periods lower than 2.5 days, our result reproduces the  Sub-Jovian desert, but we also find a scarcity of planets for masses lower than 0.8 $M_J$, and periods lower than 1 days (see e.g., ~\citealt{Ow18}).

\begin{figure}
\centering
\includegraphics[width=.42\textwidth]{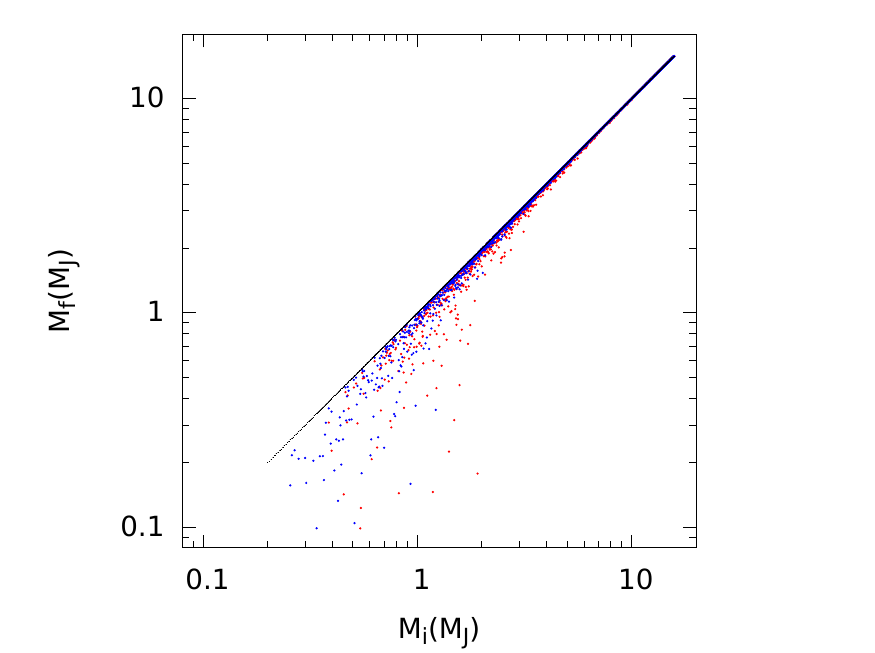}  
\caption{Final mass distribution of our planetary sample as a function of initial planetary mass. Planets in the range $T_{\rm P} = 0.8-2$~days are labelled with red (blue) dots when orbiting dG (dM) stars; black dots on the bisector are planets in the $245-418$~days period range.}
\label{fig6}
\end{figure}
Figure~\ref{fig6} shows the final vs the initial masses of close (first period bin) and far (last period bin) planets for the surviving atmospheres: low-mass close-in planets loose a substantial fraction of their initial masses, while far planets lie mainly on the bisector of the diagram (i.e. they are virtually unperturbed). We point out that we consider as vaporized, planets completely devoid of their gaseous envelopes showing thus their bare rocky core \citep{Lec04}, the so called chthonian planets, \citealt{He04}. In our simulations, we select \citet{F07} models with a constant $10~M_\oplus$ rocky core. Thus, vaporization produces rocky planets with such a final constant value for their masses. This is of course an oversimplification; e.g., \citet{lammer} argues that at distances $d_{\rm P} \sim 0.1-1$~AU, in solar-like environments the critical mass to accrete a gaseous envelope is $M_{\rm P} \sim 5-20~M_\oplus$. Thus, at shorter orbital distances, rocky planets of lower mass are expected.

We estimate the frequency of giant planets at the age of 10 Myr, per star of given spectral type, and each period bin given in \citet{F13}. To this aim we use the distributions shown in Figure~\ref{fig4} as corrections to the current planetary mass profile, this last assumed as being dominated by 4.5 Gyr old planets. In the case of dG (dM) stars, in the orbital period bin $0.8-2$~days, we find 4\% (2\%) more giant planets at early times than at the nominal final evolutionary time, and 1\% ($\sim 0.5\%$) in the  period bin $2-3.4$ days. At larger periods basically no differences occur. Such differences are small and likely unobservable, even in future much larger planetary samples. However, these differences are specifically concentrated in the $0.2-2~M_J$ range, rather than over the full mass range. For instance, in the period bin $T_{\rm P} = 0.8-2$~days, around young stars we find $\sim 36\% (\sim 20\%)$ more planets having masses ranging from Saturn to Jupiter ones. This  value increases up to a factor of at least 2 when we consider closer gaseous giants. These predictions may be confirmed or disproved by the project GAPS2 (Global Architecture of Planetary Systems), the extension of the program GAPS \citep{gaps}, for which the determination of the frequency of massive hot planets around young stars constitutes one of the main objectives. 

Finally, we note that all of the vaporized planets lie below the lifetime line in the energy diagrams shown in Figure~\ref{fig2}.

\subsection{The temporal evolution of the mass-radius relation}
\begin{figure*}
\centering
\begin{tabular}{cc}
\includegraphics[width=.35\textwidth]{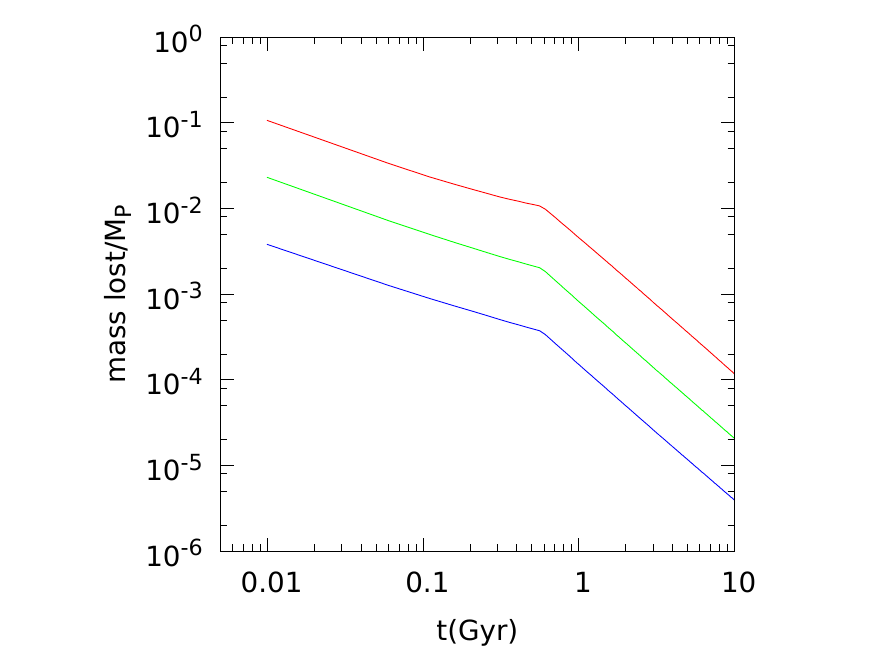} & 
\includegraphics[width=.35\textwidth]{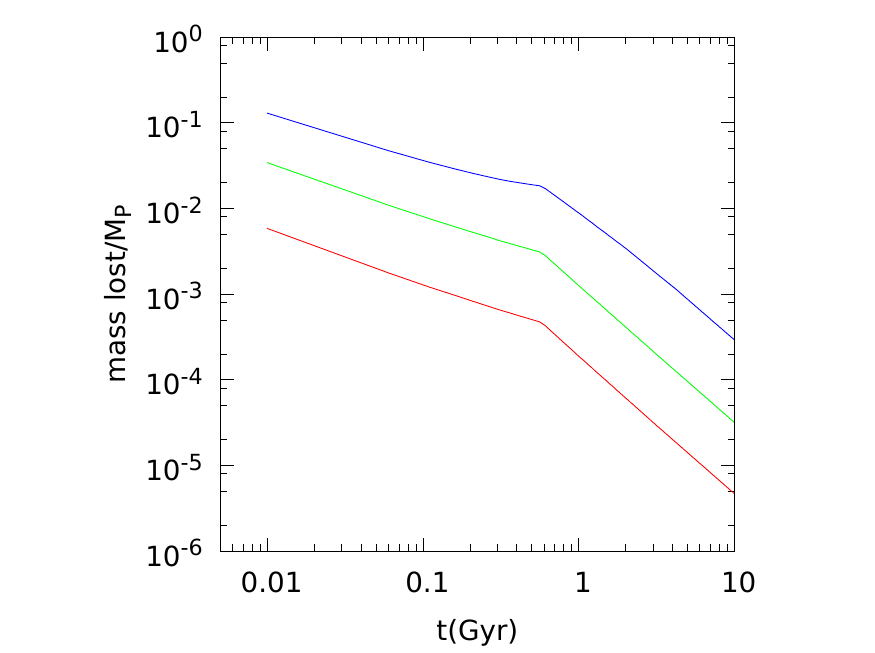} \\
\includegraphics[width=.35\textwidth]{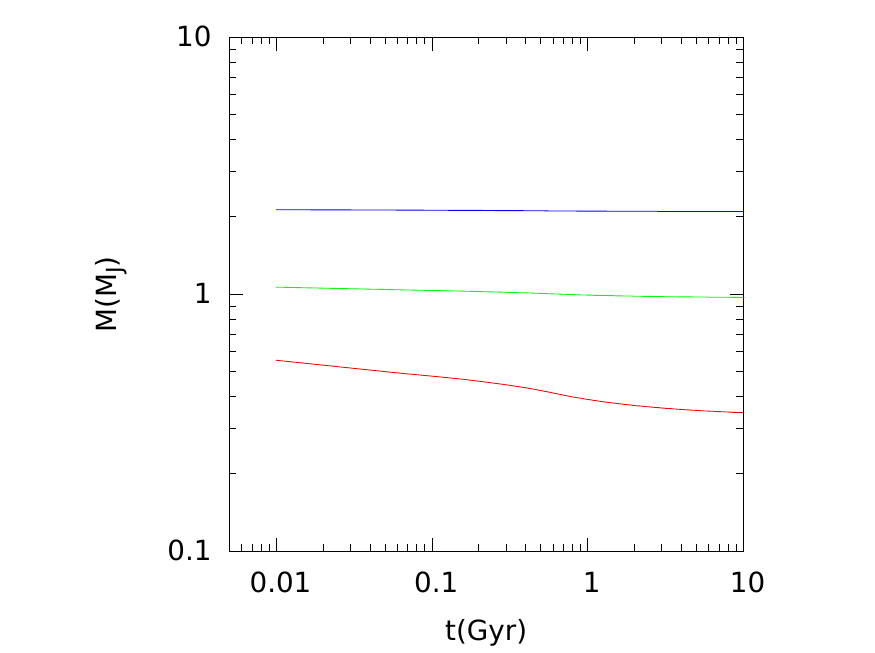} & 
\includegraphics[width=.35\textwidth]{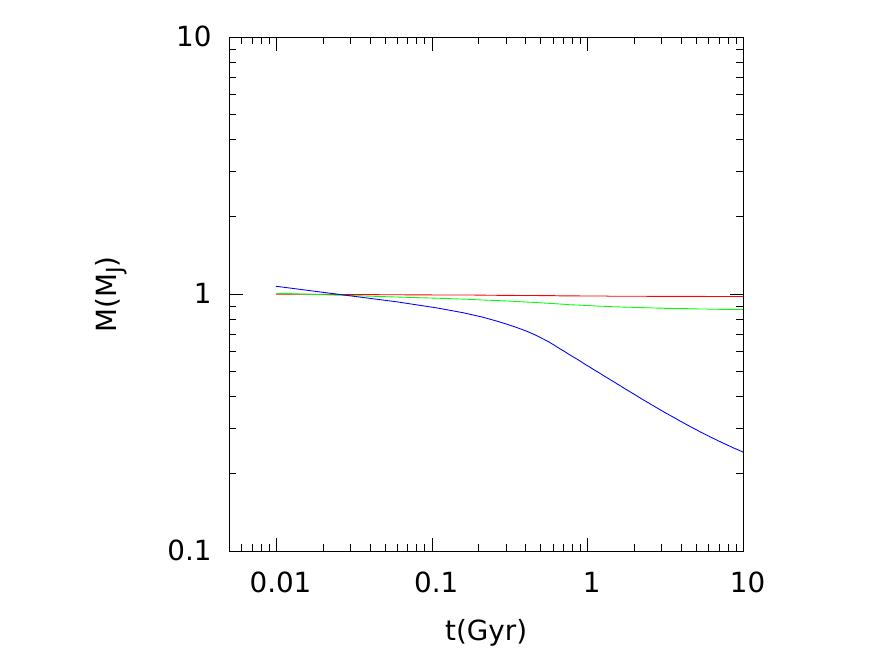} \\
\includegraphics[width=.35\textwidth]{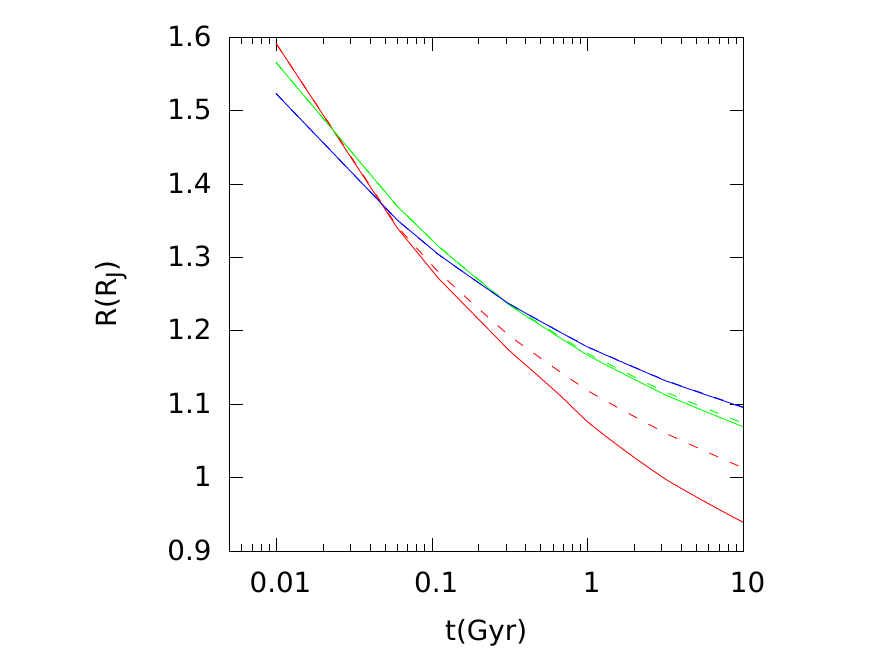} & 
\includegraphics[width=.35\textwidth]{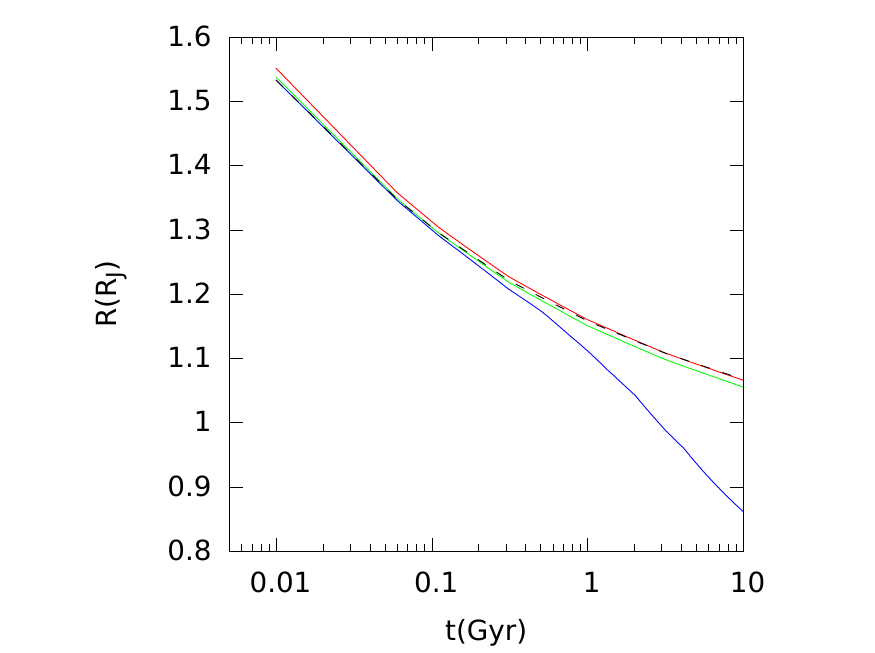} \\
\end{tabular}
\caption
{Evolving mass loss (top panels), mass (middle panels) and radius (bottom panels). In the left panels it is reported the case of three planets of different masses, $M_{\rm P} = 0.55$ (red line), 1 (green line), and 2~$M_J$ (blue line), subjected to identical irradiation $F_{\rm XUV} = 1.1 \times 10^{6}$~erg~cm$^{-2}$~s$^{-1}$; in the right panel the planetary mass is taken constant to $M_{\rm P} = 1 ~M_J$, while the initial X-ray luminosity is varied, $L^{(0)}_{\rm X} = 3 \times 10^{28}$ (red line), $3 \times 10^{29}$ (green line), $1.6 \times 10^{30}$~erg~s$^{-1}$ (blue line). The dashed  lines in the bottom left panel indicate the temporal evolution of the planetary size without evaporation (i.e. just considering gravitational shrink). In the  bottom right panel the black dashed line marks the temporal evolution of the planetary radius due to gravitational contraction. Top labels indicate the initial X-ray luminosity (left panels), and planetary mass (right panels).
}
\label{fig7}
\end{figure*}

In Figure~\ref{fig7} we highlight the effects of the high energy radiation on the temporal evolution of planetary mass and radius. We show the temporal evolution in two cases: $(i)$ three masses, $M_{\rm P} = 0.55$, 1, and 2~$M_J$, with respective orbital periods 1.7, 1.5, and 1.8 days, that receive a common initial XUV flux $F_{\rm XUV} = 1.1 \times 10^{6}$~erg~cm$^{-2}$~s$^{-1}$; $(ii)$ a single planetary mass $M_{\rm P} =1~M_J$ with a period of 1.8 days, and three initial X-ray luminosities $L^{(0)}_{\rm X} = 3 \times 10^{28}$, $3 \times 10^{29}$ and $1.6 \times 10^{30}$~erg~s$^{-1}$. All the planets are in the first period bin. From the results reported in the figure, it is clear that lighter planets lose more mass that the more massive ones (top left panel), while planets around the more bright stars lose  more mass (top right panel). An interesting point is the mass loss of $1~M_J$ planets may be significantly affected if the central star is very bright (middle right panel). Comparing The fates of irradiated planets with unperturbed ones, we find again that the most affected planets are those with lower masses, and those orbiting around brighter stars (see bottom panels).

\begin{figure*}
\centering
\includegraphics[width=.80\textwidth]{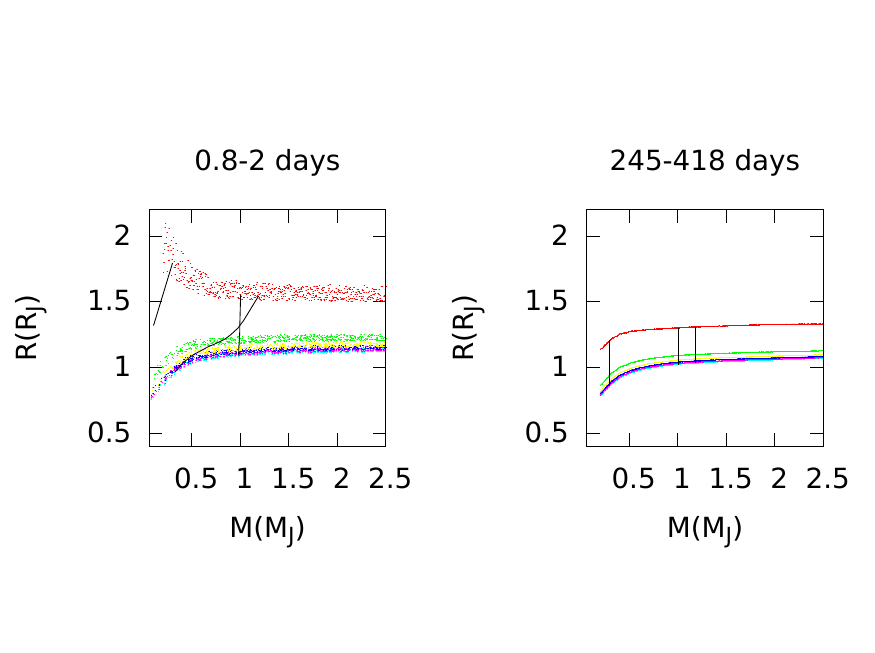}
\caption{Mass-radius relation for near (left panel) and far (right panel) exoplanets,  at different ages: 10~Myrs (red dots), 0.5~Gyrs (green dots), 1.5~Gyrs (yellow dots), 2.5~Gyrs (blue dots), 3.5 Gyrs (purple dots), and 4.5 Gyrs (light blue dots). In both panels, black lines refer to  the evolutionary paths followed by three planets with initial  masses  $M_{\rm P} = 0.6 M_J$ ($L^{0}_{\rm X}= 8.3 \times 10^{28}$~erg~s$^{-1}$), $1 M_J$ ($L^{0}_{\rm X} = 7.7 \times 10^{28}$~erg~s$^{-1}$), and $1.2 M_J$ ($L^{0}_{\rm X} = 1.6 \times 10^{30}$~erg~s$^{-1}$) nd orbital period 1.7 days}.
\label{fig8}
\end{figure*}

Finally, in Figure~\ref{fig8} we show the mass-radius evolutionary tracks of three planets orbiting around the dG star. Their initial masses are $M_{\rm P} = 0.3$, 1, and $1.2~M_J$, and we locate then in two different orbital period bins, $T_{\rm P} = 0.8-2$, and $245-418$~days. In the first period, we select planets with (almost) identical periods, i.e. $T_{\rm P} = 1.7$~days. The initial stellar X-ray luminosities are $L^{(0)}_{\rm X} = 6.12 \times 10^{28}$, $3.12 \times 10^{28}$ and $1.7 \times 10^{30}$~erg~s$^{-1}$. The first planet has been selected to explore the effects of low densities; in fact, such planet vaporizes after roughly 100~Myr, despite that the high energy flux of its parent star is relatively modest. The other two planets possessing similar initial mass and density, follow different evolutionary tracks because of the different amount of impinging stellar high energy radiation. No effects are seen for distant planets. The results shown in Figure~\ref{fig8} imply that during the first 1~Gyr, both radius and mass undergo large variations, remaining unaffected at longer times. In presence of a population of planets with an age spread, we expect  a spread in the mass-radius relation, with younger planets laying in the higher radius -  higher mass part of the diagram. From Figure~\ref{fig8} is also clear that the evolution of the planetary radius is dominated by gravitational contraction, especially during the first 0.5 Gyr. Generally, the modification to the radius in response to the atmospheric losses introduce a small correction, with the exception of planets orbiting stars with a strong high energy emission, when the mass  loss mechanism becomes important during the whole simulation. This imply a greater correction to the radius due to the mass loss (see also Figure~\ref{fig7}).

\begin{figure}
\centering
\includegraphics[width=.5\textwidth]{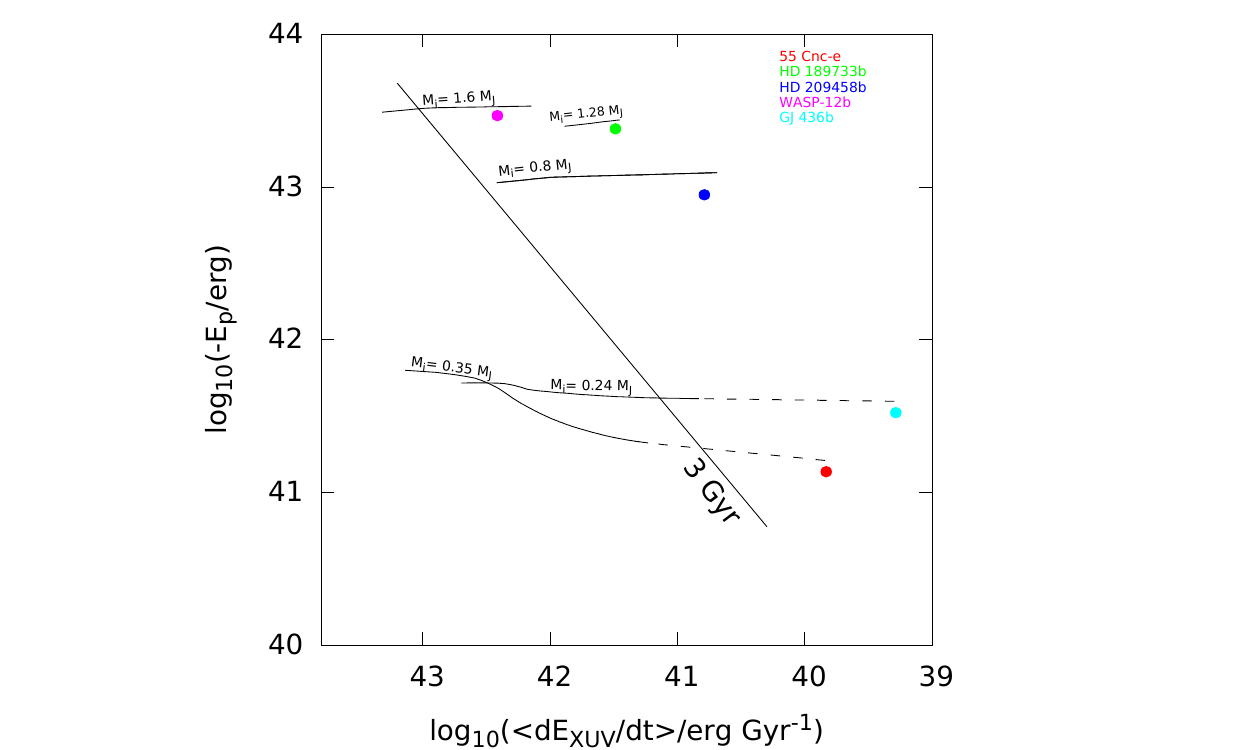}
\caption{Energy diagrams for five observed exoplanets together with the corresponding evolutionary track. Each simulation is labelled with the initial mass of the planet. The dashed part of the lines indicates that the part of the curve that is an extrapolation, as our model is unable to calculate the planetary radius of planets with masses much smaller than $0.2 M_J$.}
\label{fig9}
\end{figure}

In Figure~\ref{fig9} we show in an energy diagram the position of some well known real exoplanet, together with synthetic evolutionary tracks obtained by our modelling technique, as an indication of the evolutionary paths followed by the real planets during their lifetimes. All the planets have been assumed orbiting around G stars, although some of them are actually orbiting K stars. In Table \ref{evr} we show the mass loss rate for the planets reported in Figure~\ref{fig9}, calculated by means of  Equation~(\ref{eq:maslos3}). We also compare this results with those presents in literature, from both theoretical models and measurements.

\begin{figure}
\centering
\includegraphics[width=.5\textwidth]{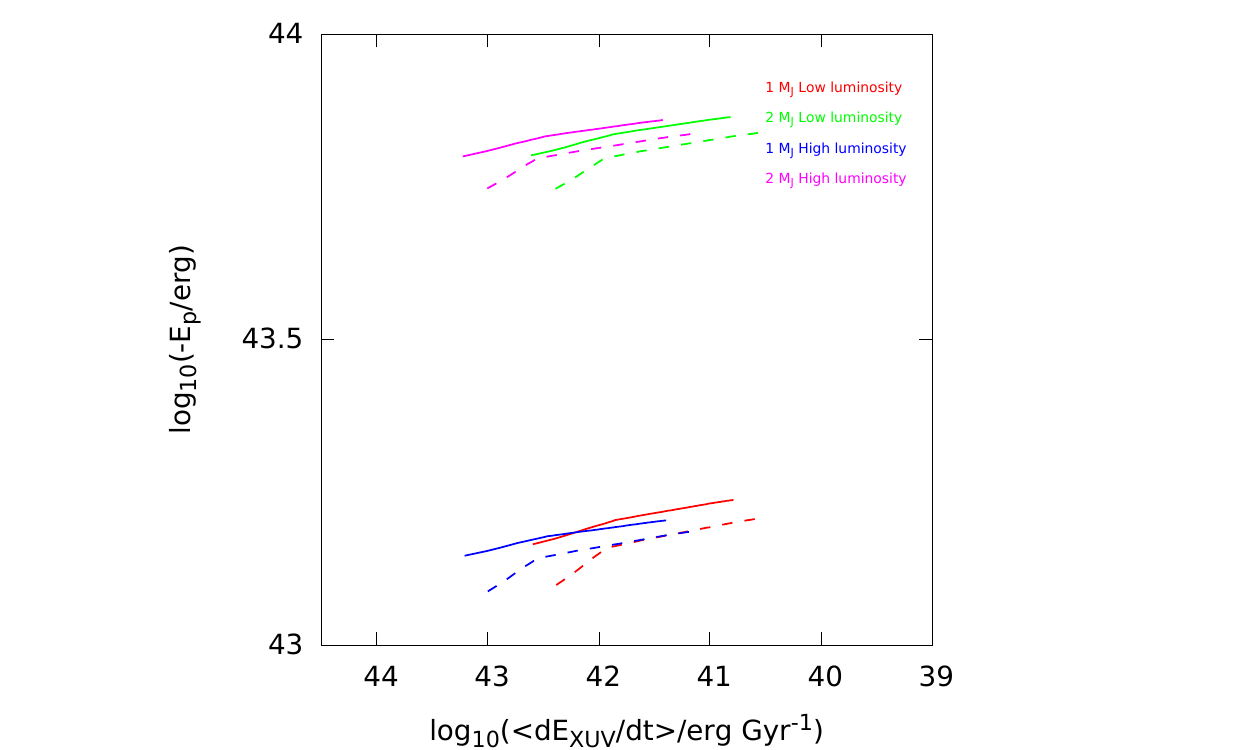}
\caption{Energy diagrams for two planets of mass $M_{\rm P} = 1$ and $2~M_J$ orbiting around dG (dashed lines) and dM (solid lines) stars of identical initial X-ray luminosities and fluxes. 
}
\label{fig10}
\end{figure} 

Since the X-ray emission of dM stars decreases at a slower rate than for dG, we compare in the energy diagram the evolutionary tracks of two planets, with mass $M_{\rm P} = 1$ and $2~M_J$, orbiting around dG and dM stars. The stars are assumed to have the same initial X-ray luminosities, namely $L^{(0)}_{\rm X} = 2 \times 10^{28}$ and $1 \times 10^{29}$~erg~s$^{-1}$, while the planets are orbiting at the same distance from the parent stars, i.e. they receive the same XUV flux. The results reported in Figure \ref{fig10} show that the evolutionary tracks tend to get closer and closer as the time increases. As a consequence the fate of an exoplanet population at the age at which planets do not evolve anymore, does not depend significantly on the stellar type.
\begin{table}
\centering
\caption{Theoretical and observational estimates of the evaporation rate $\dot{M}$ (g~s$^{-1}$) of the five planets reported in Figure~\ref{fig9}.}
\begin{tabular}{cccc}
Planet  & this work & HD modelling & observations \\
\hline 
HD 189733b & $4 \times 10^{11}$ & $4.9\times10^{9}$ $^{(1)}$ & $\sim 10^{10} $ $^{(2)}$ \\
HD 209458b & $2.5\times10^{10}$ & $1.2\times10^{10} $ $^{(1)}$ & $\sim 10^{10} $ $^{(3)}$  \\
GJ 436b &   $1.6\times10^{9}$  &  $3.95\times10^{9}$ $^{(1)}$ & $10^8-10^9$ $^{(4)}$ \\
WASP-12b$^{(5)}$ & $1.3\times10^{12}$   &  $2.7\times10^{14}$ $^{(6)}$ &   \\
55 Cnc e$^{(7)}$ &  $4.3\times10^{9}$  &  $4.2\times10^{10}$ $^{(1)}$ &  \\
\hline 
\end{tabular}
\flushleft \scriptsize
$(1)$ \citet{Kub18}; $(2)$ \citet{Lec10}; $(3)$ \citet{VM03} ; $(4)$ \citet{Eh15}; $(5)$ 
the evaporation mass loss rate is uncertain \citep{Ha18}; $(6)$ \citet{Lai10}; $(7)$ no observational rates are available.
\label{evr}
\end{table}

\subsection{Mass distribution with a constant birth rate}
In the previous subsections we have analyzed the evolution of a sample of planets having identical ages, evolving from 10~Myr to 4.5~Gyr. To provide a more realistic picture, we now consider planets at different evolutionary stages. We assume that during the simulation time ($\sim 10$~Gyr) the planet birth rate remains constant. To cover more densely the parameter space we extend the number of simulated planets to ${\cal N}_{\rm P} = 100,000$. Planet orbital periods are within the period bin $0.8-2$~days. 

\begin{figure}
\centering
\includegraphics[width=.42\textwidth]{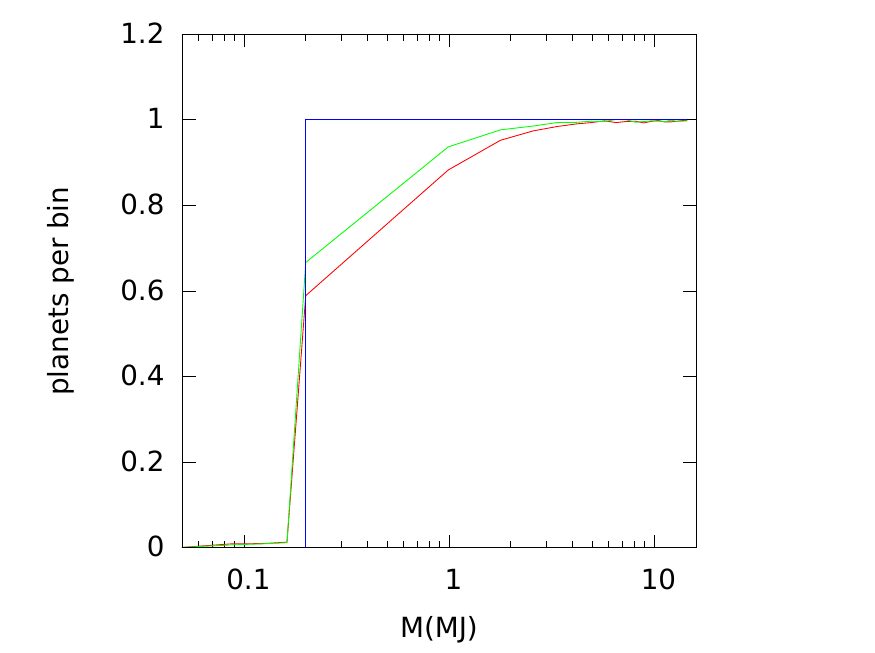} 
\caption{Number of planets per mass bin for a sample of ${\cal N}_{\rm P} = 100,000$ planets of different ages, forming at a constant (in time) birth rate (see text). The initial flat mass distribution is reported as a blue line. Red line: planets around dG stars; Green line: planets around dM stars. All the planets are in the first period bin ($0.8-2$~days).} 
\label{fig11}
\end{figure}

Far planets are unaffected by high energy radiation, their final mass distribution coinciding with the initial one. The results in terms of mass distribution are shown in Figure~\ref{fig11}. Close-in planets ~\textendash~ in particular Jovian and sub-Jovian ~\textendash~ show a final mass distribution departing significantly from the original profile, emphasizing both the efficiency and speed of radiation-induced evaporation processes. Assuming that the initial mass distribution does not depend on the orbital distance (that is not unquestionable and depends on planetary migration, e.g. \citealt{T02}), the present results suggest that planets in the $0.2 - 1~M_J$ range occur less frequently close to a XUV emitting star than far away. Such discrepancy ($\sim 41\% (33\%)$ in the case of the dG (dM) stars) in the occurring frequencies of close and far planets could be observable after correcting for the observational biases that all too often favor the detection of close-in planets. In the present calculations we find a milder difference between the occurrence frequencies of planets orbiting dG and dM stars with respect to the case of planets with identical ages (cf. Figures~\ref{fig4}, top left panel, and \ref{fig11}). This is consistent with the results shown in Figure~\ref{fig10}.

\section{Discussion and conclusions} \label{disc}
The present model simulates the hydrodynamic escape of planetary atmospheres embedded in intense XUV radiation fluxes from parent stars. The model is based on the energy-limited approximation. During the simulations the planetary radius changes in response to gravitational contraction and photo-evaporation. The relevant factors in the simulation are the planetary distance and the magnitude of stellar XUV luminosity. This latter parameter depends on time as low mass stars are known to have evolving radiation environments. Mass escape depends implicitly, but significantly on the stellar mass, as this last one is crucial in determining the planet distance (in a fixed period bin), and thus the flux impinging at the top of the atmosphere, and the variation in time of the strength of XUV radiation (see equations \ref{lxtime} and \ref{EUV}).

The general results from this work are as follows.
\begin{itemize}
\item[$(i)$] a significant fraction (4\% around dG stars and 2\% around dM stars) of low mass Jupiter-like planets orbiting with periods lower than 6 days vaporize during the first billion years, and 1\% around dG stars and 2\% around dM stars of planets loose a relevant part of their atmospheres. 
\item[$(ii)$] the planetary initial mass profile is significantly distorted in stellar radiation environments dominated by high energy photons; in particular, the frequency of occurrence of planets in the $0.2-2~M_J$ mass range around young star can be considerably greater (20\% and 10\% in the case of the dG  and dM stars, respectively) than the planetary occurring frequency around older counterparts.
\item[$(iii)$] We find that the mass loss rates around dG and dM stars differ significantly at early times, but such differences decrease with time.
\end{itemize}

The results of the present calculation provide a correction factor, that if applied to the current frequency of exoplanets, allows to retrieve the frequency around young stars.

In conclusion, XUV stellar illumination may have important consequences on the mass distribution of close-in planets, heating the outer layers of a planet’s envelope and driving mass loss. However, other mechanisms may have significant impact on the planetary mass distribution. Moreover, all these processes might operate simultaneously. A possible diagnostics may be the mass-loss dependence on stellar mass \citep{FP18}, because of its tight link to the high-energy luminosity, in particular during the pre-main sequence phase when the emission is dominated by intense flares (e.g. \citealt{Fa05}).

\begin{acknowledgements}
We acknowledge support from INAF through the “Progetto Premiale: A Way to Other Worlds” of the Italian Ministry of Education. We would like to thank the anonymous referee for her/his comments, suggestions, and criticisms that contributed to improve the manuscript. 
\end{acknowledgements}

\end{document}